\pdfoutput=1
\documentclass[11pt]{article}
\usepackage[T1]{fontenc}
\usepackage[utf8]{inputenc}
\usepackage{lmodern}
\usepackage[margin=0.82in]{geometry}
\usepackage{amsmath,amssymb,bm}
\usepackage{graphicx}
\usepackage{booktabs,tabularx,array}
\usepackage{microtype}
\usepackage{xurl}
\usepackage[hidelinks]{hyperref}
\usepackage{placeins}
\usepackage{float}
\usepackage{enumitem}
\newcolumntype{Y}{>{\raggedright\arraybackslash}X}
\newcolumntype{C}{>{\centering\arraybackslash}X}
\newcolumntype{R}{>{\raggedleft\arraybackslash}X}
\title{Fragmentation of Virtual Orbitals for Quantum Computing: Reducing Qubit Requirements through Many-Body Expansion}
\author{Federico Zahariev$^{1,2}$, Vassiliki-Alexandra Glezakou$^{3}$, and Mark S. Gordon$^{1,2}$}
\date{}
\begin{document}
\maketitle
\begin{center}
\small $^{1}$Department of Chemistry, Iowa State University, Ames, Iowa 50011, USA\\$^{2}$Ames National Laboratory, Ames, Iowa 50011, USA\\$^{3}$Chemical Sciences Division, Oak Ridge National Laboratory, Oak Ridge, Tennessee 37831, USA
\end{center}
\vspace{0.5em}
\begin{abstract}
We introduce virtual-orbital fragmentation (Q-FVO), a systematic method for reducing the largest active space in correlated quantum-chemistry calculations. The complete occupied space is retained, the localized virtual space is partitioned into chemically motivated fragments, and the correlation energy is recovered through an inclusion--exclusion many-body expansion. Across six molecular benchmarks, the largest one-body Q-FVO calculations reduce the qubit requirement by 46--66\%, while two-body calculations reduce it by approximately 31--42\% relative to the corresponding unfragmented spaces. Two-body expansions recover most of the correlation energy, with errors of 0.9--7.5 kcal mol$^{-1}$; three-body expansions are sub-kcal mol$^{-1}$ for all CCSD tests and remain below 2 kcal mol$^{-1}$ at CCSD(T). Illustrative statevector UCCSD calculations also reduce implementation-reported circuit depth while retaining sub-kcal mol$^{-1}$ accuracy. Q-FVO can be nested inside Q-EFMO real-space fragmentation and, in turn, the resulting cluster can be embedded in a Q-EFP environment. The hierarchy therefore reduces quantum-resource growth along three complementary dimensions: environment, molecular fragments, and virtual-orbital space.
\end{abstract}
\vspace{0.5em}
\section{Introduction}
Near-term quantum-chemistry calculations are constrained by qubit count, circuit depth, measurement cost, and noise \cite{cao2019,mcardle2020,bauer2020,peruzzo2014,mcclean2016,preskill2018}. Under common fermion-to-qubit mappings, each spin orbital requires one qubit \cite{jordan1928,bravyi2002}. Virtual orbitals often dominate this count, particularly in polarized and augmented basis sets, even though their primary role is to represent electron correlation. Reducing the virtual space in a controlled and systematically improvable manner is therefore a direct route to smaller quantum subproblems.

Existing strategies include chemically chosen active spaces, tensor-network or selected-configuration decompositions, adaptive ansatz construction, and measurement or error-mitigation methods \cite{gunst2018,tubman2020,tilly2022,grimsley2019,ryabinkin2018,huggins2021,zhao2021,temme2017,li2017}. Spatial fragmentation methods such as FMO and EFMO instead divide a large system into monomers and near-field dimers and recover the total energy through a many-body expansion \cite{kitaura1999,steinmann2010,pruitt2013,pruitt2014,gordon2012}. These real-space approaches reduce total problem size and expose parallelism. Quantum Monte Carlo/EFMO studies have shown that many-body-expansion fragmentation retains high accuracy even for demanding correlated calculations, with errors below 1\% for water clusters at a cutoff of 1.4 van der Waals radii and correlation-energy errors below 2 kcal mol$^{-1}$ when fragmenting across covalent bonds in glycine oligomers, dipeptide formation, silica-based materials, and polyalanine chains \cite{zahariev2019,bertoni2016,fmo_book,zahariev2021}. However, the active orbital space of an individual fragment may still be too large for the available register.

We introduce virtual-orbital fragmentation (Q-FVO), an orthogonal decomposition that partitions the localized virtual space while retaining the complete occupied space in every calculation. Correlation is reconstructed through an inclusion--exclusion many-body expansion over virtual fragments. The hierarchy can be converged by adding higher-body combinations, and each reduced-space calculation is independent. Q-FVO can also be nested inside EFMO, creating a two-level Q-EFMO/Q-FVO decomposition in real space and orbital space.

The physical motivation is locality and redundancy in the virtual space. Excitations from occupied orbitals preferentially couple to nearby virtual functions, especially after localization, while large polarized and augmented bases contain multiple virtual orbitals describing similar spatial and energy regions. Retaining every occupied orbital avoids artificial changes to bonding and electron counting; only the virtual correlation space is partitioned. This makes Q-FVO complementary to, rather than a replacement for, frozen cores, frozen natural orbitals, or chemically selected active spaces.

The present benchmarks evaluate qubit reduction and many-body convergence for six molecular systems, illustrate UCCSD statevector circuit-depth reductions for two active spaces, and test the hierarchical workflow on four molecular clusters. Because circuit depth and cluster active spaces are implementation-specific quantities, those results are presented as proof-of-workflow resource illustrations rather than definitive hardware benchmarks.

\section{Theory and computational methods}
\subsection{Virtual-orbital many-body expansion}
\begin{figure}[H]
\centering
\includegraphics[width=0.98\textwidth]{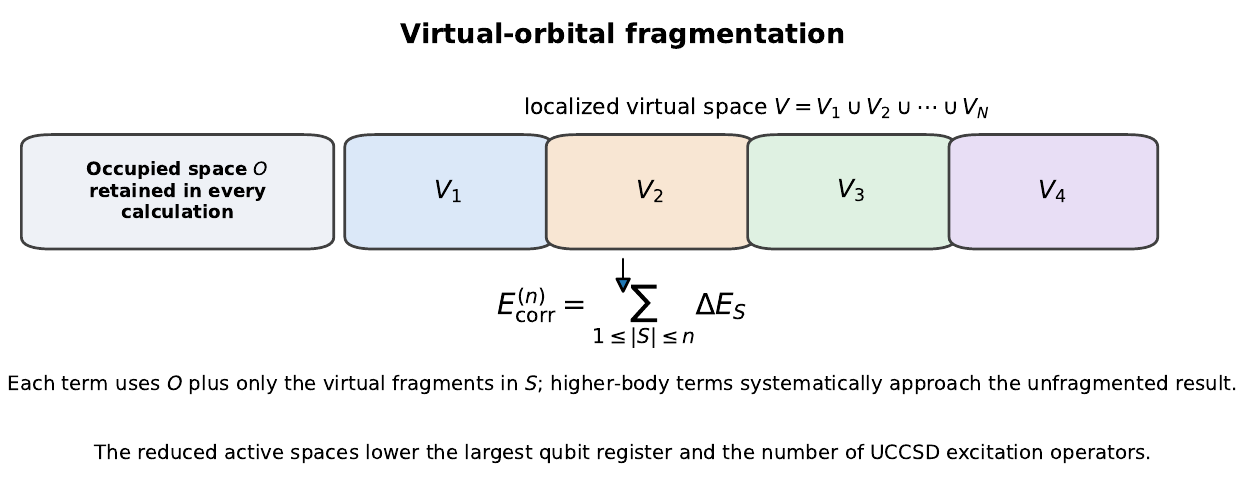}
\caption{Virtual-orbital fragmentation. The complete occupied space is retained, the localized virtual space is divided into fragments, and correlation is recovered through a many-body expansion over subsets of virtual fragments. Each reduced active space requires fewer qubits than the unfragmented calculation.}
\label{fig:fvo_schematic}
\end{figure}

Let the full orbital space be the union of an occupied space $O$ and a localized virtual space $V$. Q-FVO partitions $V$ into nonoverlapping subsets $V_1,\ldots,V_N$ while retaining all orbitals in $O$ in every correlated calculation (Figure~\ref{fig:fvo_schematic}). For a subset $S\subseteq\{1,\ldots,N\}$, define $V_S=\bigcup_{i\in S}V_i$ and evaluate the correlation energy $E_{\mathrm{corr}}(O\cup V_S)$.

The $n$-body approximation is

\begin{equation}
E_{\mathrm{corr}}^{\mathrm{Q\mbox{-}Q-FVO}(n)}=\sum_{\substack{S\subseteq\{1,\ldots,N\}\\1\le |S|\le n}}\Delta E_S,
\label{eq:1}
\end{equation}
with increments defined recursively by

\begin{equation}
\Delta E_S=E_{\mathrm{corr}}(O\cup V_S)-\sum_{T\subset S}\Delta E_T.
\label{eq:2}
\end{equation}
For example, $\Delta E_i=E_{\mathrm{corr}}(O\cup V_i)$ and $\Delta E_{ij}=E_{\mathrm{corr}}(O\cup V_i\cup V_j)-\Delta E_i-\Delta E_j$. The recursive definition prevents double counting and guarantees that the full result is recovered when all $N$-body terms are included. Smaller virtual fragments provide larger qubit savings but require more subset calculations and may shift correlation to higher orders.

\subsection{Fragment construction and integration with VQE}
Virtual orbitals are localized with Boys or Pipek--Mezey procedures \cite{boys1966,pipek1989} and assigned to molecular units, atomic centers, or chemically recognizable groups. The intended partition captures short-range occupied--virtual correlation in one-body terms and interfragment coupling in two- and higher-body increments. In the present tests, fragments containing roughly 20--40\% of the virtual space provide a useful compromise between register size and expansion order.

Each Q-FVO subset defines an independent VQE calculation containing the full occupied space and only the selected virtual fragments. With UCCSD, reducing the number of virtual orbitals decreases the number of qubits, excitation amplitudes, entangling operations, and Hamiltonian terms. Q-FVO is compatible with frozen cores, frozen natural orbitals, and chemically selected active spaces; it is a decomposition layer rather than a replacement for those approximations.

For multi-molecular systems, EFMO can first partition the cluster into monomers and near-field dimers \cite{jensen1996,gordon2007,gordon2001,zahariev2019,bertoni2016,fmo_book,zahariev2021}. Q-FVO is then applied independently within each quantum monomer or dimer calculation. The resulting Q-EFMO/Q-FVO hierarchy reduces the chemical problem in real space and the register size in virtual-orbital space.

\subsection{Relation to Q-EFP and Q-EFMO}
The three companion methods act at distinct levels. Q-EFP replaces a large solvent or ionic environment with a first-principles embedding potential, Q-EFMO partitions the remaining molecular cluster into real-space monomers and near-field dimers, and Q-FVO reduces the virtual space within each quantum monomer or dimer. Their combination does not change the occupied-space chemical partition used by Q-FVO; it simply applies the virtual-space expansion to each selected Q-EFMO subproblem while the outer EFP environment remains classical.

\subsection{Computational details}
All classical electronic-structure calculations were performed with GAMESS \cite{gamess1993,gordon2005}. Geometries were optimized with B3LYP/6-31G(d) unless otherwise specified. Q-FVO tests used CCSD and CCSD(T) correlation energies with the basis sets listed in Table~\ref{tab:fvo_systems}. Virtual orbitals were localized by Boys or Pipek--Mezey procedures as indicated for each benchmark. VQE tests used statevector UCCSD with COBYLA parameter optimization and an energy convergence threshold of $10^{-6}$ hartree. The optimizer, localization choice, and active-orbital lists for each benchmark are specified in the corresponding input files. No hardware noise, finite-shot sampling, or error mitigation was included.

\section{Results and discussion}
\subsection{Qubit reduction across six molecular systems}
The virtual space comprises 62--90\% of the spatial orbital count in the six test systems (Table~\ref{tab:fvo_systems}), so it dominates the unfragmented register. The largest cases, H$_2$O$_2$/aug-cc-pVDZ and NH$_3$/aug-cc-pVDZ, require 128 and 100 qubits, respectively, before virtual-space fragmentation.

\begin{figure}[H]
\centering
\includegraphics[width=0.84\textwidth]{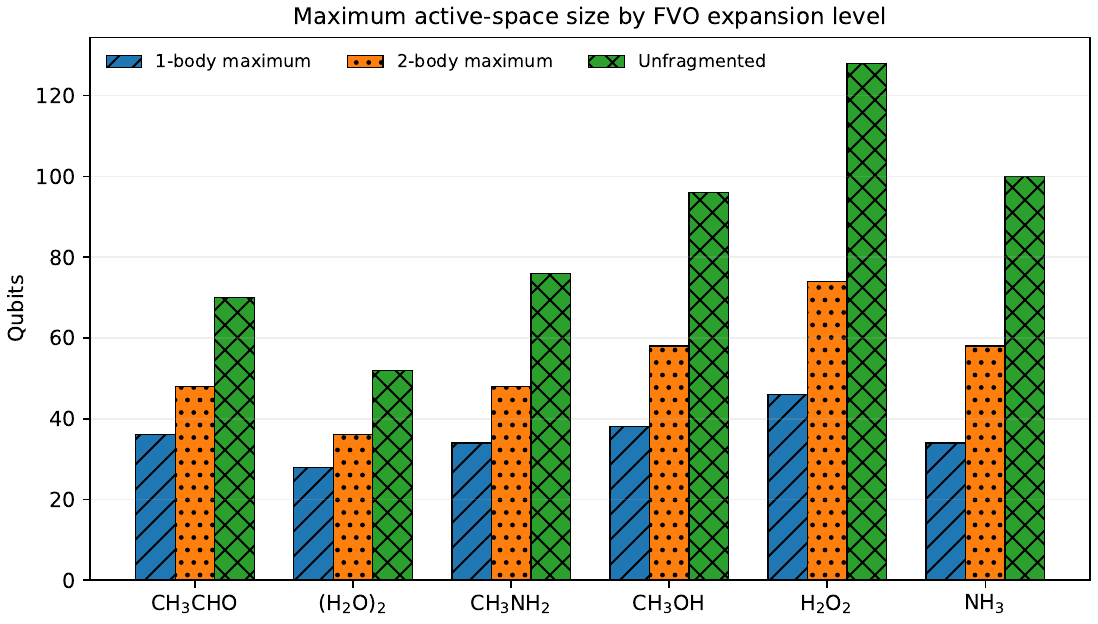}
\caption{Maximum active-space qubit count for one-body and two-body Q-FVO calculations compared with the unfragmented space. The largest savings occur when the virtual space is a large fraction of the total orbital space.}
\label{fig:fvo_qubits}
\end{figure}

\begin{table}[H]
\centering
\small
\caption{Orbital and qubit counts for the six molecular benchmarks.}
\label{tab:fvo_systems}
\begin{tabularx}{\textwidth}{@{}YYCCCC@{}}
\toprule
Molecule & Basis & Spatial orbitals & Occupied & Virtual & Full qubits \\
\midrule
CH$_3$CHO & 6-31G & $35$ & $12$ & $23$ & $70$ \\
(H$_2$O)$_2$ & 6-31G & $26$ & $10$ & $16$ & $52$ \\
CH$_3$NH$_2$ & 6-31G(d) & $38$ & $9$ & $29$ & $76$ \\
CH$_3$OH & cc-pVDZ & $48$ & $9$ & $39$ & $96$ \\
H$_2$O$_2$ & aug-cc-pVDZ & $64$ & $9$ & $55$ & $128$ \\
NH$_3$ & aug-cc-pVDZ & $50$ & $5$ & $45$ & $100$ \\
\bottomrule
\end{tabularx}
\vspace{0.35em}\parbox{0.98\textwidth}{\footnotesize Qubit counts assume one qubit per spin orbital and no symmetry tapering.}
\end{table}
Table~\ref{tab:fvo_qubits} and Figure~\ref{fig:fvo_qubits} show the maximum register required at the one- and two-body levels. One-body calculations reduce the maximum by 46--66\%, and two-body calculations reduce it by approximately 31--42\%. For H$_2$O$_2$, the largest two-body calculation requires 74 rather than 128 qubits; for NH$_3$, it requires 58 rather than 100.

\begin{table}[H]
\centering
\small
\caption{Maximum qubit requirement at different Q-FVO expansion levels.}
\label{tab:fvo_qubits}
\begin{tabularx}{\textwidth}{@{}YCCC@{}}
\toprule
Molecule & 1-body maximum & 2-body maximum & Unfragmented \\
\midrule
CH$_3$CHO & $36$ & $48$ & $70$ \\
(H$_2$O)$_2$ & $28$ & $36$ & $52$ \\
CH$_3$NH$_2$ & $34$ & $48$ & $76$ \\
CH$_3$OH & $38$ & $58$ & $96$ \\
H$_2$O$_2$ & $46$ & $74$ & $128$ \\
NH$_3$ & $34$ & $58$ & $100$ \\
\bottomrule
\end{tabularx}
\end{table}
\subsection{Accuracy of the many-body expansion}
The one-body expansion captures substantial correlation but leaves absolute errors of 57--147 kcal mol$^{-1}$ (Table~\ref{tab:fvo_errors}). Adding all two-body increments reduces the errors to 0.9--7.5 kcal mol$^{-1}$ and recovers approximately 96--99.5\% of the full correlation energy. At the three-body level, all CCSD errors are below $0.53$ kcal mol$^{-1}$, and all CCSD(T) errors are below $1.69$ kcal mol$^{-1}$. The similar order-by-order trends at CCSD and CCSD(T) indicate that the partition error is not strongly specific to one of these two correlation models.

\begin{table}[H]
\centering
\small
\caption{Absolute Q-FVO errors in kcal mol$^{-1}$ relative to the full unfragmented calculation.}
\label{tab:fvo_errors}
\begin{tabularx}{\textwidth}{@{}YYCCC@{}}
\toprule
Molecule & Method & 1-body & 2-body & 3-body \\
\midrule
CH$_3$CHO & CCSD & $66.648$ & $2.950$ & $0.070$ \\
 & CCSD(T) & $70.553$ & $1.919$ & $0.152$ \\
(H$_2$O)$_2$ & CCSD & $56.754$ & $2.526$ & $0.001$ \\
 & CCSD(T) & $57.713$ & $2.298$ & $0.087$ \\
CH$_3$NH$_2$ & CCSD & $86.409$ & $0.938$ & $0.460$ \\
 & CCSD(T) & $90.135$ & $0.974$ & $0.257$ \\
CH$_3$OH & CCSD & $89.087$ & $5.136$ & $0.490$ \\
 & CCSD(T) & $93.395$ & $7.511$ & $0.632$ \\
H$_2$O$_2$ & CCSD & $140.081$ & $2.458$ & $0.524$ \\
 & CCSD(T) & $147.430$ & $7.042$ & $1.685$ \\
NH$_3$ & CCSD & $59.338$ & $2.966$ & $0.213$ \\
 & CCSD(T) & $62.406$ & $4.698$ & $0.672$ \\
\bottomrule
\end{tabularx}
\vspace{0.35em}\parbox{0.98\textwidth}{\footnotesize Values are errors in total correlation energy at the indicated many-body order.}
\end{table}
\begin{figure}[H]
\centering
\includegraphics[width=0.80\textwidth]{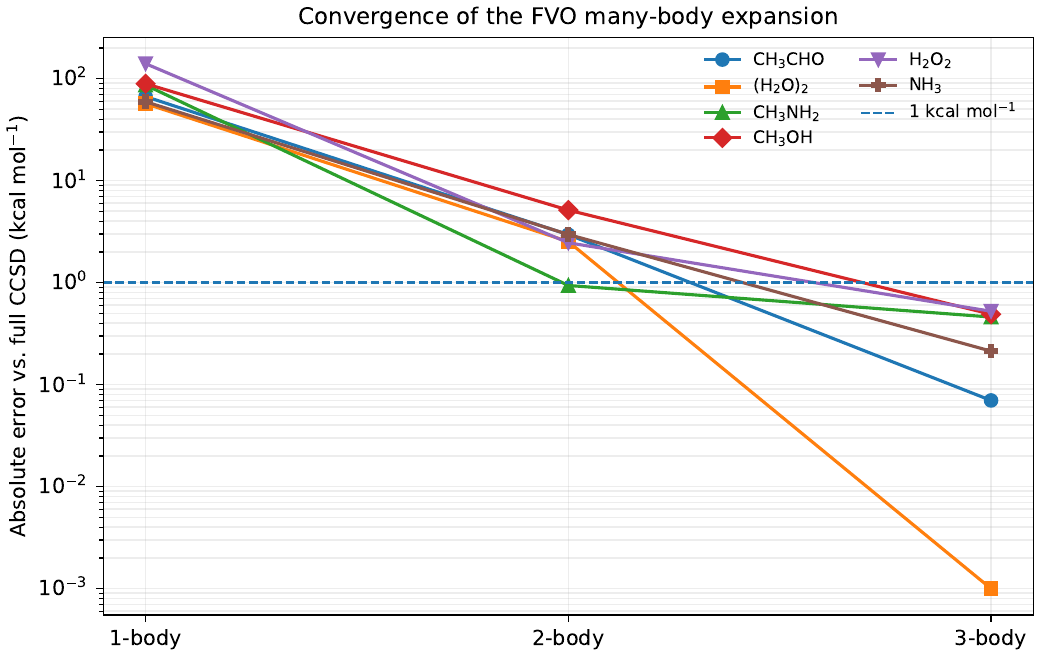}
\caption{Convergence of the CCSD Q-FVO expansion. The logarithmic vertical axis shows that the error falls by one to several orders of magnitude between the one- and three-body levels. The dashed line marks 1 kcal mol$^{-1}$.}
\label{fig:fvo_convergence}
\end{figure}
\subsection{Illustrative VQE circuit reductions}
\begin{figure}[H]
\centering
\includegraphics[width=0.74\textwidth]{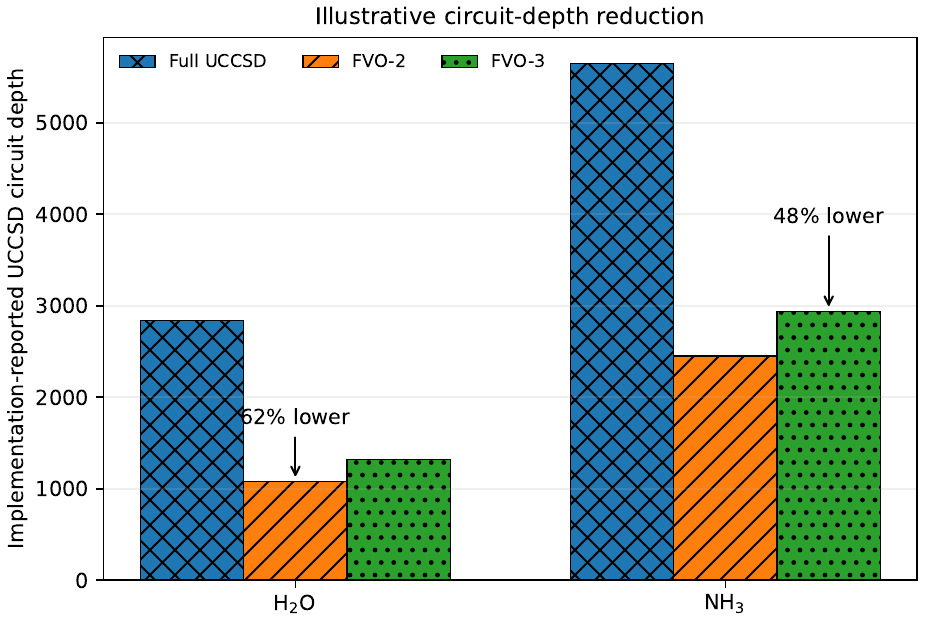}
\caption{Illustrative UCCSD circuit-depth reductions for the H$_2$O and NH$_3$ active spaces in Table~\ref{tab:fvo_vqe}. Percentages compare the indicated Q-FVO calculation with the corresponding unfragmented circuit.}
\label{fig:fvo_depth}
\end{figure}

Statevector UCCSD circuit builds for two separately selected active spaces show that Q-FVO can reduce both qubit count and implementation-reported ansatz depth (Table~\ref{tab:fvo_vqe} and Figure~\ref{fig:fvo_depth}). For the H$_2$O test, Q-FVO-2 reduces the register from 52 to 32 qubits and the listed circuit depth by 62\%, with an energy error of $0.48$ kcal mol$^{-1}$. For the NH$_3$ test, Q-FVO-3 uses 56 rather than 92 qubits and reduces the listed depth by 48\%, with an error of $0.31$ kcal mol$^{-1}$.

The active spaces used for these circuit constructions are implementation-specific and are not identical to the orbital-count survey in Table~\ref{tab:fvo_systems}. Circuit depth also depends on fermion mapping, gate set, compilation level, connectivity, and cancellation rules. The values should therefore be interpreted as relative reductions within one workflow, not as portable hardware-depth predictions.

\begin{table}[H]
\centering
\small
\caption{Illustrative statevector UCCSD resource and accuracy data for Q-FVO active spaces.}
\label{tab:fvo_vqe}
\begin{tabularx}{\textwidth}{@{}YYCCC@{}}
\toprule
System & Method & Qubits & Circuit depth & Energy error (kcal mol$^{-1}$) \\
\midrule
H$_2$O & Full UCCSD & $52$ & $2840$ & -- \\
 & Q-FVO-2 & $32$ & $1080$ & $0.48$ \\
 & Q-FVO-3 & $36$ & $1320$ & $0.42$ \\
NH$_3$ & Full UCCSD & $92$ & $5650$ & -- \\
 & Q-FVO-2 & $48$ & $2450$ & $0.58$ \\
 & Q-FVO-3 & $56$ & $2940$ & $0.31$ \\
\bottomrule
\end{tabularx}
\vspace{0.35em}\parbox{0.98\textwidth}{\footnotesize Circuit depth is implementation reported and depends on the compilation convention; the active spaces are distinct from Table~\ref{tab:fvo_systems}.}
\end{table}
\subsection{Hierarchical Q-EFMO/Q-FVO tests}
\begin{figure}[H]
\centering
\includegraphics[width=0.85\textwidth]{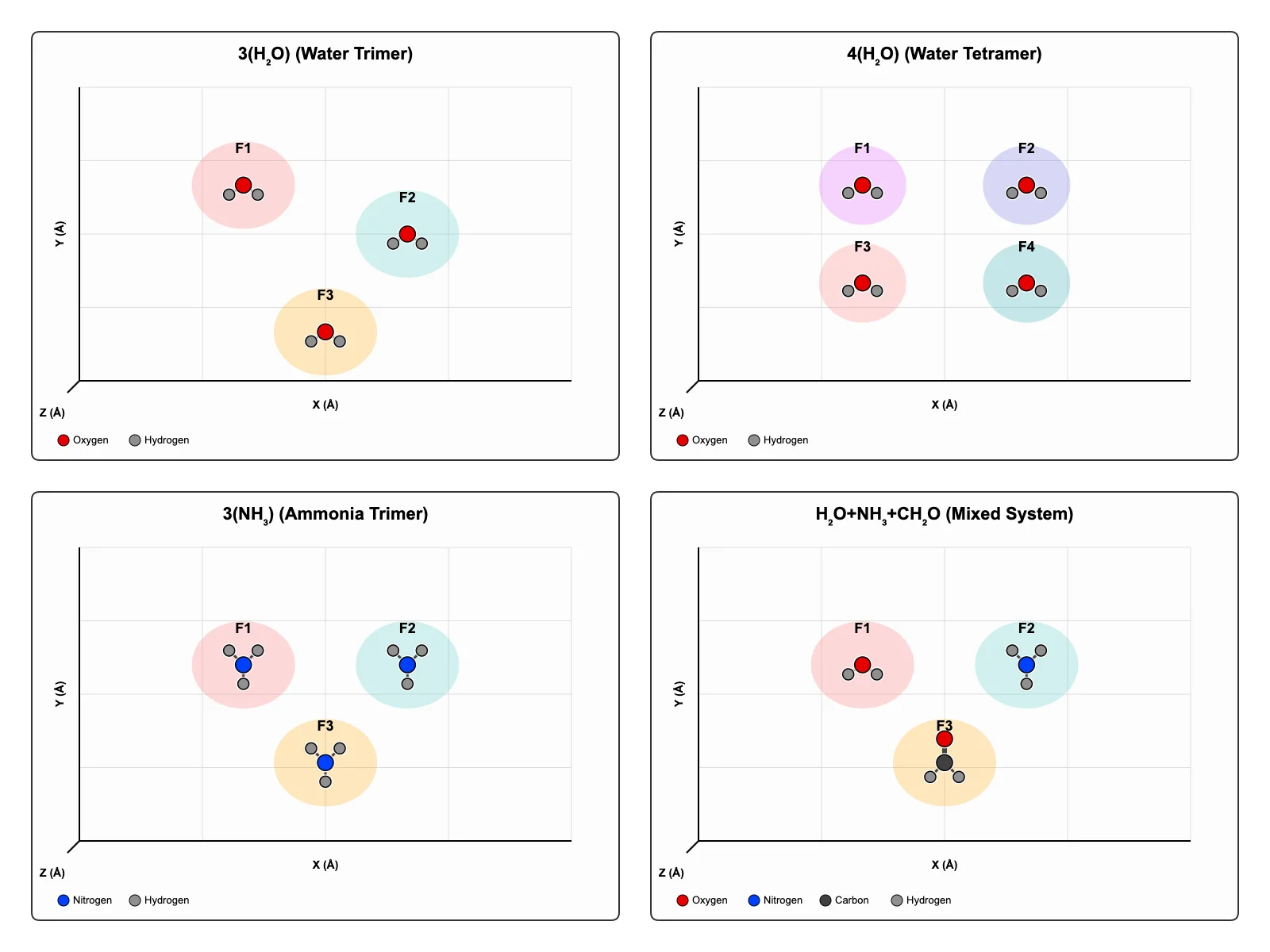}
\caption{Spatial fragmentation of the four molecular clusters used in the hierarchical Q-EFMO/Q-FVO tests. Each cluster is first decomposed into molecular monomers at the EFMO level, and the virtual space of each fragment calculation is then partitioned by Q-FVO.}
\label{fig:fvo_clusters}
\end{figure}

The hierarchical workflow was exercised for a water trimer, water tetramer, ammonia trimer, and a mixed H$_2$O + NH$_3$ + CH$_2$O cluster (Figure~\ref{fig:fvo_clusters} and Table~\ref{tab:fvo_hier}).

 In the water trimer, classical CCSD(T)-Q-FVO-2 differs from the CCSD(T)-EFMO reference by $0.982$ kcal mol$^{-1}$. The UCCSD-based cluster entries differ by $5.9$--$13.4$ kcal mol$^{-1}$ from CCSD(T)-EFMO. These deviations combine fragmentation, orbital-space, and correlation-model differences; in particular, UCCSD is not expected to reproduce perturbative triples exactly. The cluster calculations therefore demonstrate data flow and resource decomposition rather than a clean Q-FVO accuracy benchmark.

\begin{table}[H]
\centering
\small
\caption{Hierarchical Q-EFMO/Q-FVO workflow tests in the 6-31G basis.}
\label{tab:fvo_hier}
\begin{tabularx}{\textwidth}{@{}YYYCC@{}}
\toprule
System & Method & Q-FVO level / localization & $E_{\mathrm{corr}}$ (hartree) & Error vs. CCSD(T)-EFMO \\
\midrule
Water trimer & CCSD(T)-EFMO reference & -- & $-0.410$ & reference \\
 & Classical CCSD(T) & 0 / none & $-0.410$ & $0.000$ \\
 & VQE-UCCSD & 0 / none & $-0.399$ & $7.060$ \\
 & Classical CCSD(T) & 2 / Pipek--Mezey & $-0.408$ & $0.982$ \\
 & VQE-UCCSD & 2 / Pipek--Mezey & $-0.400$ & $6.230$ \\
Water tetramer & CCSD(T)-EFMO reference & -- & $-0.543$ & reference \\
 & VQE-UCCSD & 0 / none & $-0.521$ & $13.306$ \\
 & VQE-UCCSD & 2 / Pipek--Mezey & $-0.521$ & $13.370$ \\
Ammonia trimer & CCSD(T)-EFMO reference & -- & $-0.405$ & reference \\
 & VQE-UCCSD & 0 / Boys & $-0.395$ & $6.313$ \\
 & VQE-UCCSD & 2 / Pipek--Mezey & $-0.392$ & $8.434$ \\
Mixed cluster & CCSD(T)-EFMO reference & -- & $-0.504$ & reference \\
 & VQE-UCCSD & 0 / Boys & $-0.494$ & $5.911$ \\
 & VQE-UCCSD & 2 / Pipek--Mezey & $-0.492$ & $7.544$ \\
\bottomrule
\end{tabularx}
\vspace{0.35em}\parbox{0.98\textwidth}{\footnotesize Errors are in kcal mol$^{-1}$ relative to the CCSD(T)-EFMO value. VQE rows use UCCSD fragment calculations and include a correlation-model mismatch with the CCSD(T) reference.}
\end{table}
\subsection{Relation to other orbital-reduction strategies}
Q-FVO differs from a fixed active-space truncation because omitted virtual sectors are recovered systematically through higher-body increments. Frozen natural orbitals rank virtuals by occupation, whereas Q-FVO groups localized virtuals and retains intergroup couplings order by order. The methods are complementary: a frozen core, natural-orbital threshold, or chemically chosen active space can first reduce the global orbital set, after which Q-FVO can partition the remaining virtuals. The appropriate hierarchy will depend on the balance among register size, number of subset calculations, and target accuracy.

\section{Conclusions}
Virtual-orbital fragmentation reduces the maximum active space of correlated quantum-chemistry calculations by partitioning localized virtual orbitals and recovering correlation through an inclusion--exclusion many-body expansion. Across six molecular tests, one-body calculations reduce the largest register by 46--66\% and two-body calculations by approximately 31--42\%. Three-body Q-FVO reaches sub-kcal mol$^{-1}$ CCSD accuracy for every test and remains below $2$ kcal mol$^{-1}$ at CCSD(T). Illustrative UCCSD statevector builds also show substantial reductions in register size and circuit depth. Nested inside EFMO, Q-FVO provides a second, orbital-space decomposition layer. Several directions follow naturally: adaptive selection of virtual fragments from importance measures, integration with error-mitigation strategies, extension to excited states through equation-of-motion or time-dependent formulations, combination with frozen-natural-orbital or active-space preselection, and demonstration on quantum hardware. Reproducible release of fragment definitions, active spaces, and compilation settings is the prerequisite step toward evaluating the method on larger bases and hardware.

\section{Acknowledgments}
This material is based upon work supported by the National Science Foundation under Grant No. OSI-2435255. This manuscript has been authored by UT-Battelle, LLC, under Contract No. DE-AC05-00OR22725 with the U.S. Department of Energy. This work was also supported by the U.S. Department of Energy, Office of Science, through Ames National Laboratory under Contract No. DE-AC02-07CH11358. The authors acknowledge partial support by ORNL LDRD and VSO programs. The U.S. Government retains, and the publisher by accepting the article for publication acknowledges, a nonexclusive, paid-up, irrevocable, worldwide license to publish or reproduce the published form of this manuscript, or allow others to do so, for U.S. Government purposes. This research used resources of the Oak Ridge Leadership Computing Facility (Frontier; Director's Discretionary allocation CHM238) and the National Energy Research Scientific Computing Center under award m4621 (ERCAP0036406). Computational resources were also provided by Iowa State University.

\section{Data and code availability}
GAMESS is available through its project distribution. Input geometries, localized-orbital assignments, virtual-fragment membership, active-space definitions, raw subset energies, and circuit-compilation settings are available from the corresponding author upon reasonable request. These data are required to reproduce Tables~\ref{tab:fvo_qubits}--\ref{tab:fvo_hier}.

\FloatBarrier

\end{document}